\documentclass[article,twocolumn,prb,showpacs,superscriptaddress]{revtex4-1}

\usepackage{graphicx}
\usepackage{amssymb}
\usepackage{bm}
\usepackage{amsmath}
\usepackage{xfrac}
\usepackage{color}
\usepackage{titlesec}

\usepackage[normalem]{ulem}
\usepackage[utf8]{inputenc}
\usepackage[T1]{fontenc}

\usepackage[colorlinks=True,
                 linkcolor=black,
	        citecolor=blue,
	        urlcolor=blue]{hyperref}

\newcommand\3{$_3$}

\newcommand{\dqmp}{Department of Quantum Matter Physics, University of Geneva, 24 Quai Ernest Ansermet, CH-1211 Geneva, Switzerland}
\newcommand{\gap}{Group of Applied Physics, University of Geneva, 24 Quai
Ernest Ansermet, CH-1211 Geneva, Switzerland}	
\newcommand{\RCFM}{Research Center for Functional Materials,National Institute for Materials Science, 1-1 Namiki, Tsukuba, 305-0044, Japan}
\newcommand{\ICMN}{International Center for Materials Nanoarchitectonics,National Institute for Materials Science, 1-1 Namiki, Tsukuba, 305-0044, Japan}
\newcommand{\MOE}{MOE Key Laboratory for Nonequilibrium Synthesis and Modulation of Condensed Matter, School of Physics, Xi’an Jiaotong University, Xi’an,710049, China}
\newcommand{\unimore}{Dipartimento di Scienze Fisiche, Informatiche e Matematiche, University of Modena and Reggio Emilia, IT-41125 Modena, Italy}

\makeatletter
\g@addto@macro\bfseries{\boldmath}
\makeatother

\begin{document}
\title{Magnetization dependent tunneling conductance of ferromagnetic barriers}
\author{Zhe Wang}
\affiliation{\MOE}
\affiliation{\dqmp}
\author{Ignacio Guti\'errez-Lezama}
\affiliation{\dqmp}
\affiliation{\gap}
\author{Dumitru Dumcenco}
\affiliation{\dqmp}
\author{Nicolas Ubrig}
\affiliation{\dqmp}
\affiliation{\gap}
\author{Takashi Taniguchi}
\affiliation{\ICMN}
\author{Kenji Watanabe}
\affiliation{\RCFM}
\author{Enrico Giannini}
\affiliation{\dqmp}
\author{Marco Gibertini}
\affiliation{\unimore}
\affiliation{\dqmp}
\author{Alberto F.\ Morpurgo}
\affiliation{\dqmp}
\affiliation{\gap}


\maketitle

{\bf
Recent experiments on  van der Waals antiferrmagnets such as CrI\3, CrCl\3 and MnPS\3 have shown that using atomically thin layers as tunnel barriers and measuring the temperature ($T$) and magnetic field ($H$) dependence of the conductance allows their magnetic phase diagram to be mapped. In contrast, barriers made of CrBr\3 --the sole van der Waals ferromagnet investigated in this way-- were  found to exhibit small and featureless magnetoconductance, seemingly carrying little information about magnetism. Here we show that --despite these early results-- the conductance of CrBr\3 tunnel barriers does provide detailed information about the magnetic state of atomically thin CrBr\3 crystals for $T$ both above and below the Curie temperature ($T_C = 32$ K). Our analysis establishes that the tunneling conductance depends on $H$ and $T$ exclusively through the magnetization $M(H,T)$, over the entire temperature range investigated (2-50 K). The phenomenon is reproduced in detail by the spin-dependent Fowler-Nordheim model for tunneling, and is a direct manifestation of the spin splitting of the CrBr\3 conduction band. These findings demonstrate that the investigation of magnetism by tunneling conductance measurements is not limited to antiferromagnets, but can also be applied to ferromagnetic materials.
}

Probing magnetism in atomically thin van der Waals crystals is challenging because most experimental methods commonly employed to study bulk compounds are not sufficiently sensitive to detect any magnetic signal from such a small amount of material\cite{Burch2018,Gong2019,Marco2019,Mak2019,avsar2020,Huang2020}. Recently, it has been shown that magnetic phase boundaries --and even the complete magnetic phase diagram-- of insulating atomically thin magnets can be detected by using them as tunnel barriers, and measuring their temperature-dependent magnetoconductance\cite{Song2018Science,Klein2018Science,Wang2018,wang2019NN,cai2019NL,klein2019NP,Kim2019PNAS,Kim2019NL,Long2020NL}. The sensitivity of the tunneling magnetoconductance to magnetism originates from the dependence of the tunneling probability on the magnetic state\cite{moodera2007review}. As $H$ and $T$ are varied across a magnetic transition, the alignment of the spins in the barrier changes sharply, and so does the tunneling probability of electrons with different spin orientations. The net result is an equally sharp change in the measured conductance that can be traced to identify the phase boundary.

\begin{figure*}
\centering
\includegraphics[width =0.8\linewidth]{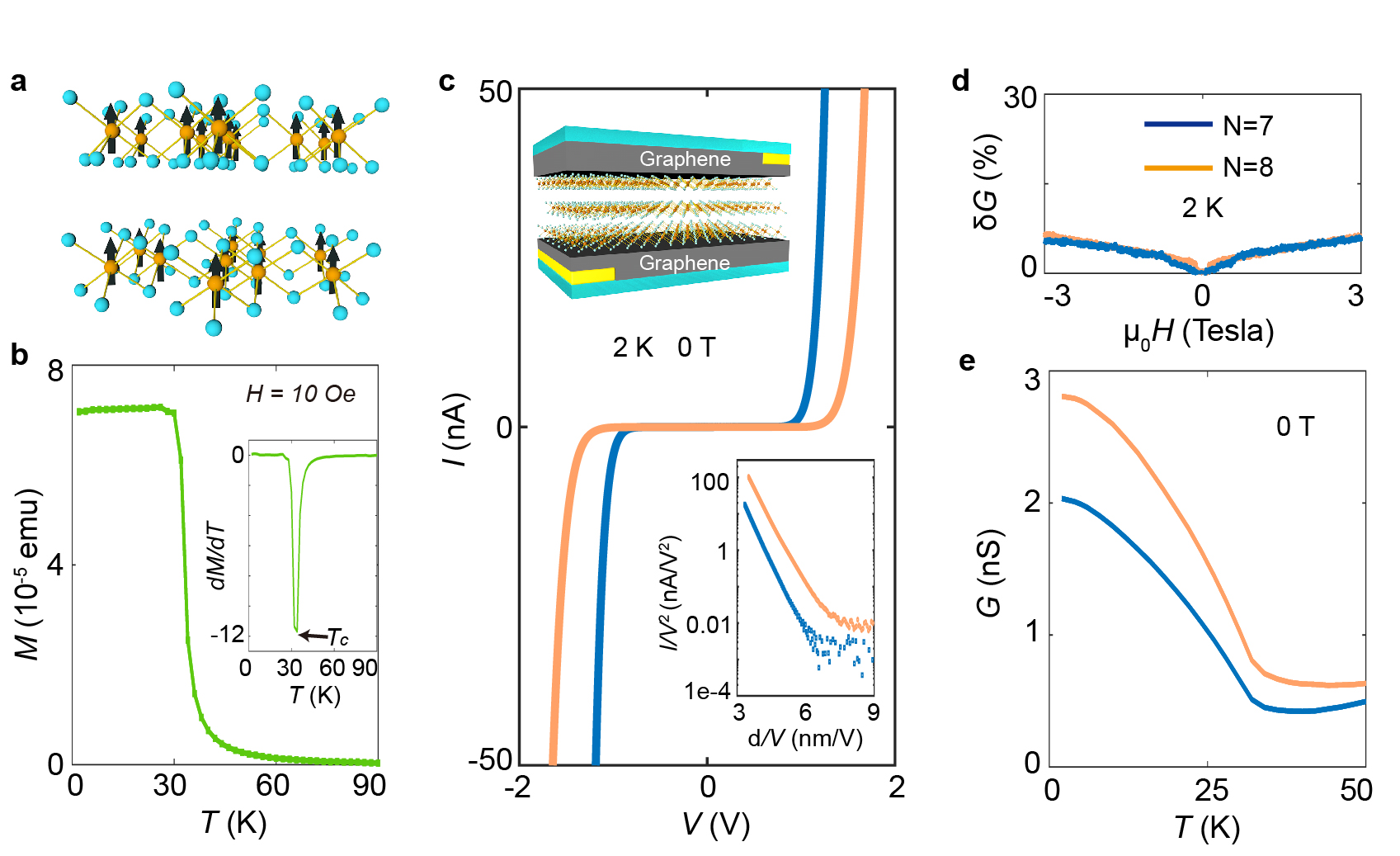}
\caption{\textbf{Tunneling conductance of CrBr\3 multilayers}. {\bf a}: Schematics of the crystal structure of CrBr\3: the orange balls represent the Cr atoms with the associated spins pointing perpendicularly to the layers; the light blue balls represent the Br atoms. {\bf b}: Temperature dependence of the magnetization measured on bulk CrBr\3 crystals with a magnetic field of 1 mT applied perpendicular to the layers. The inset shows the plot of $dM/dT$, with a sharp minimum close to 32 K, corresponding to the Curie temperature of CrBr$_3$. {\bf c}: Tunneling current as a function of applied voltage measured on a  7 (blue curve) and 8 (orange curve) layer CrBr\3 device at $T=2$ K (curves of the same color  in panels {\bf d} and {\bf e} represent data measured on the same devices). The up left insert is a cartoon representation of our hBN-encapsulated graphene/CrBr\3/graphene tunnel junction devices. The down right insert shows that --for sufficiently large applied bias $V$--  $\log(I/V^2)$ is linearly  proportional to $d/V$ ($d$ is the thickness of CrBr\3),  as expectd in the  Fowler-Nordheim tunneling regime. {\bf d}: Magnetoconductance $\delta G(H,T) \equiv [G(H,T)-G (0,T)]/G(0,T)$ measured at $T=2$ K, on the  7 and 8 layer devices.   {\bf e}: Temperature dependence of the zero-field tunneling conductance of the 7 and 8 layer devices, exhibiting an increase of approximately 300 \%, as $T$ is lowered from  $T_c$ to 2 K. Data in panels {\bf d} and {\bf e} have been acquired with an applied bias voltage $V =$ 1 V and $V= 1.4$ V for the 7 and 8 layer devices, respectively.} 
\label{fig:tun}
\end{figure*}

These conclusions have been drawn from experiments on different antiferromagnetic insulators (CrI\3\cite{Song2018Science,Klein2018Science,Wang2018}, CrCl\3\cite{wang2019NN,cai2019NL,klein2019NP,Kim2019PNAS} and MnPS\3\cite{Long2020NL}), and it is not at all clear that  the technique can be equally effective to probe ferromagnetic insulators, since in that case  no magnetic phase boundaries are present below the Curie temperature. Indeed, tunneling conductance measurements on ferromagnetic CrBr\3 barriers have shown only an extremely small and featureless low-temperature tunneling magnetoconductance\cite{Ghazaryan2018,Kim2019PNAS,Kim2019NL}, and no pronounced effect was observed, which could be  related to magnetism without microscopic modeling. In contrast to this early results, here we show directly from experimental data that the temperature and magnetic field dependence of the tunneling conductance of  ferromagnetic CrBr\3 barriers is entirely determined by the magnetization of the material, and can be used to extract detailed, quantitative information both above and below the Curie temperature.

CrBr\3 is a van der Waals layered material that --irrespective of thickness (i.e., from bulk down to monolayer)--  exhibits a transition to  a ferromagnetic state with an easy axis perpendicular to the layers\cite{Tsubokawa1960,Jennings1965PR,Ho1969,Neutron1971,Kim2019NE,Chen2019Science,zhang2019NL,Jin2020NM,Sun2021} (see Fig.~\ref{fig:tun}a, Supplementary Note 1 and Supplementary Fig.~1). Bulk magnetization measurements in Fig.~\ref{fig:tun}b show that the Curie temperature of our crystals is $T_C\simeq 32$~K, with a  saturation magnetization  corresponding to a magnetic moment of $3\mu_B$ per chromium atom, in line with previous reports\cite{Tsubokawa1960,Jennings1965PR,Ho1969,Neutron1971}. Single crystals are exfoliated into thin layers and used to nano-fabricate hBN-encapsulated graphite/CrBr\3/graphite tunnel junctions inside a glove box (see inset of Fig.~\ref{fig:tun}c for a scheme, Supplementary Fig.~2 for an optical image of the device, and Methods for detailed information about device assembly). Fig.~\ref{fig:tun}c presents the  current-voltage ($I$-$V$) characteristics for two representative devices with different thickness $d$ (corresponding to $N=7$ and 8 layers), showing typical tunneling transport at low temperatures and the scaling behavior predicted by the Fowler-Nordheim tunneling formula (i.e., $\log(I/V^2)\propto d/V$)\cite{FN1928,FN1969}.  As shown in Fig.~\ref{fig:tun}d,  application of an external magnetic field up to 3~T at $T=2$~K causes only minor (<6\%) and featureless variations  in the conductance $G=I/V$. This is consistent with previous reports\cite{Kim2019PNAS,Kim2019NL} and expected for a ferromagnetic semiconductor, in which --at low temperature-- the spins are spontaneously fully polarized already in the absence of an applied magnetic field.

Despite the negligible low-temperature magnetoconductance, Fig.~\ref{fig:tun}e shows that the conductance measured at zero applied magnetic field $G(H=0, T)$ increases by a factor of three as $T$ is lowered from the Curie temperature $T_C=32$~K down to 2~K. As the conductance is virtually temperature independent above $T_C$ (for $32 <T< 50$ K), we infer directly from the experimental data that the conductance increase is due to  CrBr\3  entering the ferromagnetic state. This observation implies that magnetism does influence the electrical conductance of the tunnel barriers and that the effect is sizable: a threefold increase in conductance is comparable to the magnetoresistance of  most common magnetic tunnel junctions (i.e., tunneling spin valve devices\cite{vzutic2004}) and of CrCl\3 antiferromagnetic tunnel  barriers\cite{cai2019NL,klein2019NP,wang2019NN,Kim2019PNAS,Kim2019NL}. 

\begin{figure*}
\centering
\includegraphics[width =0.9\linewidth]{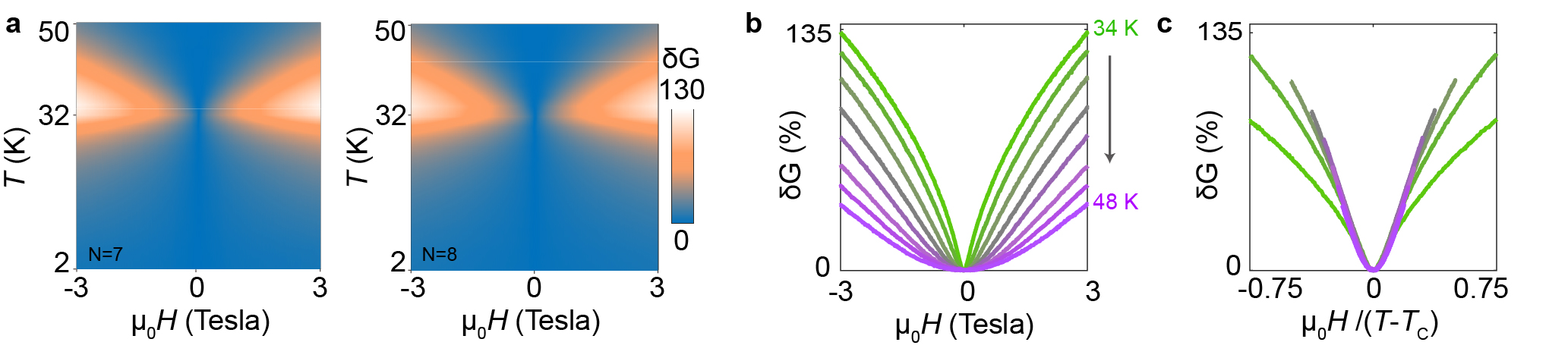}
\caption{\textbf{Temperature evolution of the tunneling magnetoconductance}. {\bf a}: Color plot of tunneling magnetoconductance for the 7 (left) and 8 layer (right) device as a function of applied magnetic field $\mu_0 H$ and temperature $T$. At fixed applied  field, the magnetoconductance is  maximum close to  $T_c=32$ K. {\bf b}: Magnetoconductance of 7 layer device for $T > T_c$, as $T$ is varied from 34 to 48 K, in 2 K steps.  {\bf c}: when plotted as a function of $\mu_0 H/(T-T_C)$, the magnetoconductance curves shown in panel {\bf b} collapse on top of each other at small field, indicating that in the linear regime the magnetocoductance $\delta G(H,T)$ depends on $H$ and $T$ only through the magnetization $M(H,T)$. In panels {\bf b}  and {\bf c}, curves of the same color correspond to  measurements done at the same temperature.}
\label{fig:MG}
\end{figure*}

This observation motivates us to look more in detail at the temperature-dependent magnetoconductance of CrBr\3 barriers, $\delta G(H,T) \equiv [G(H,T)-G (0,T)]/G(0,T)$. The full dependence of  $\delta G(H,T) $ on $T$ and $H$ is shown in Fig.~\ref{fig:MG}a for both the 7- and 8-layer CrBr\3 devices, with the two of them exhibiting identical behavior. In both devices, $\delta G(H,T)$  is positive and peaks at $T = T_C$ irrespective of the applied magnetic field $H$. The positive magnetoconductance can be understood, since at high $T$ the application of a magnetic field does lead to a better alignment of the spins in CrBr\3. Similarly to what happens in CrI\3 and CrCl\3 tunnel barriers, a better spin alignment enhances the tunneling probability, causing the conductance to increase. We note that when $T$ approaches $T_C$ from above --coming from the paramagnetic state of CrBr\3-- the magnetic field required to increase the conductance systematically decreases, indicating that the spin susceptibility $\chi$ is enhanced. This trend is reminiscent of the behavior expected from the critical fluctuations in the neighborhood of the ferromagnetic transition\cite{blundell2003magnetism}.

The idea that the magnetoconductance for $T>T_C$ probes the fluctuations of the spins in the critical regime of the paramagnetic state can be tested quantitatively if we recall that in a mean-field description of this regime, the linear spin polarizability $\chi \propto 1/(T-T_C)$ as $T$ approaches $T_C$\cite{blundell2003magnetism}.  We can then check whether the conductance depends on the magnetic field induced spin polarization, or equivalently on the magnetization (whose mean-field expression is given by the Curie-Weiss law, $M \propto \mu_0 H/(T-T_C)$), by simply plotting  $\delta G(H,T)$   as a function of  $\mu_0 H/(T-T_C)$ for any $T>T_C$ (see Fig.~\ref{fig:MG}b). For sufficiently small $\mu_0 H/(T-T_C)$ all curves  indeed  collapse on top of each other (Fig.~\ref{fig:MG}c), irrespective of the temperature at which they are measured, confirming that in the linear regime the field-induced increase of the conductance is determined by the net spin polarization (i.e., by the field-induced magnetization).

\begin{figure*}
\centering
\includegraphics[width =0.8\linewidth]{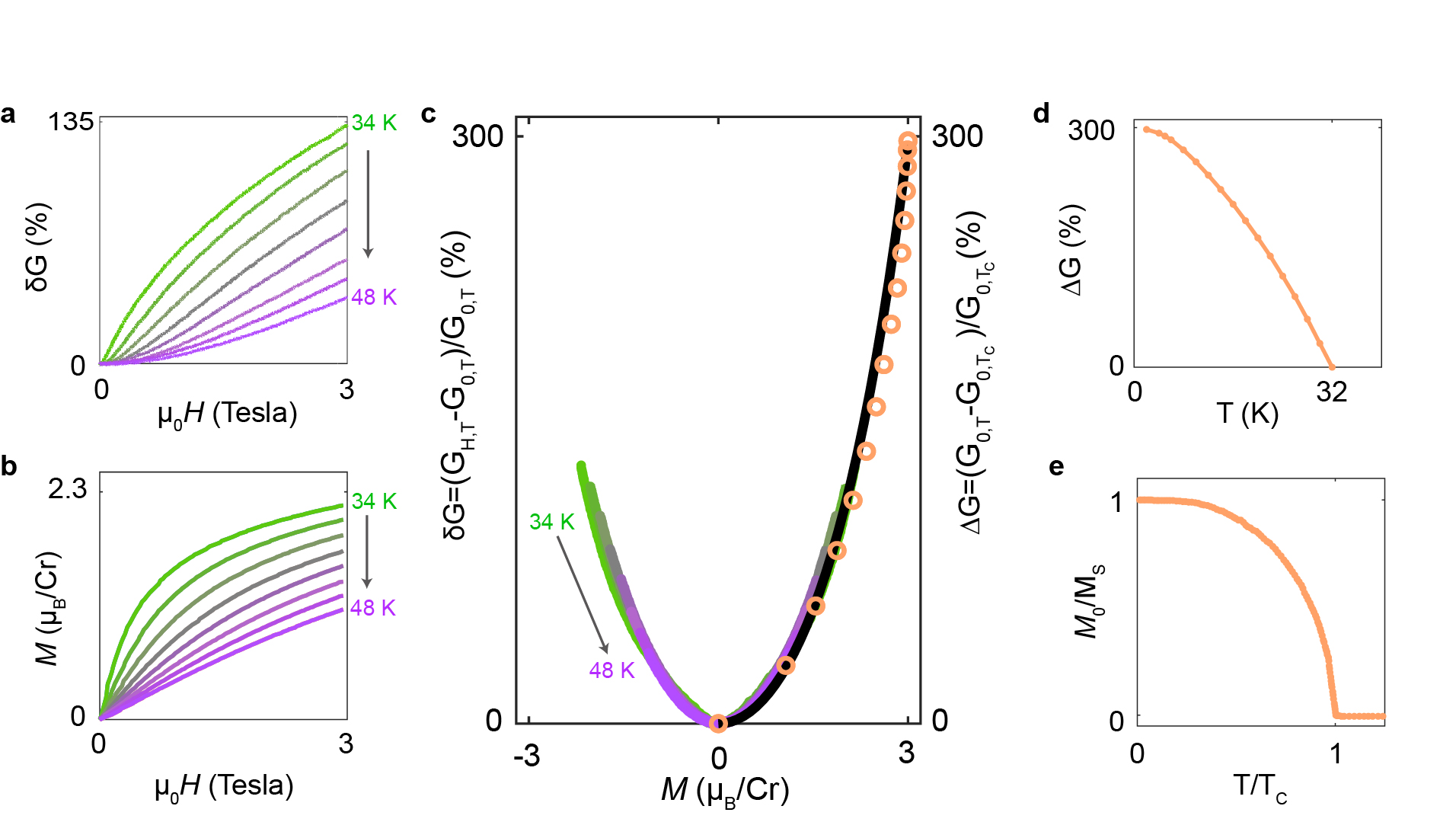}
\caption{\textbf{Magnetization dependence of tunneling magnetoconductance}. {\bf a}: Tunneling magnetoconductance measured for  $T>T_c$, as $T$ is varied from 34 K to 48 K in 2 K steps  (curves of the same color  in panels {\bf b} and {\bf c} represent measurements taken at the same temperature). {\bf b}: Magnetic field dependence of the magnetization measured on bulk CrBr\3 crystals. {\bf c}: Magnetization dependence of tunneling magneto-conductance. Colored lines represent the  magnetoconductance measured at different   $T>T_c$, plotted as a function of the bulk magnetization measured at the same temperature. The orange open circles represent the relative change in conductance due to the increase in the spontaneous magnetization of CrBr$_3$, measured for different $T<T_c$, obtained from the data shown in panels {\bf d} and {\bf e}. All data collapse on  top of each other, indicating that  the conductance is a function of the  magnetization, i.e. it depends on $H$ and $T$ exclusively through $M(H,T)$, throughout the entire $T$ range investigated (i.e., from well below to well above $T_C$). The black line is a fit based on the expression obtained from the spin-dependent Fowler-Nordheim tunneling model, under the assumption that the splitting between the spin up and down bands is proportional to the magnetization (see main text for details). {\bf d}: Temperature dependence of the relative conductance increase as $T$ is lowered below $T_C$, in the ferromagnetic state of CrBr$_3$. {\bf e}: Spontaneous magnetization of CrBr$_3$ calculated by XXZ model with anisotropic exchange interactions that --as shown in Ref. \onlinecite{Kim2019NE}-- accurately reproduces the measured magnetization of atomically thin CrBr\3 crystals . }
\label{fig:scaling}
\end{figure*}

The relation between magnetoconductance and magnetization can be tested beyond the linear regime, by using the magnetization $M(H,T)$ measured on bulk crystals (Fig.~\ref{fig:scaling}b) to re-plot the magnetoconductance of our tunnel barriers $\delta G(H, T)$ (Fig.~\ref{fig:scaling}a) as a function of $M$. The result is shown in Fig.~\ref{fig:scaling}c, with curves of different colors representing magnetoconductance measurements done at different temperatures. When plotted as a function of $M$ all curves collapse on top of each other throughout the entire range of $H$ and $T>T_C$ investigated. We can therefore conclude directly from the data that  the magnetoconductance $\delta G$ depends on $H$ and $T$ only through the magnetization $M(H,T)$ even well outside the linear regime. That is:  for $T>T_C$, $\delta G(H, T) = \delta G (M (H,T))$ .

To extend our analysis from the paramagnetic state to $T<T_C$, when the CrBr\3 barriers are ferromagnetic, we look at the temperature dependence of the conductance measured at zero applied field. To this end, we consider the quantity $\Delta G(H,T) \equiv [G(0,T)-G(0,T_C)]/G(0,T_C)$, i.e., the relative increase in conductance observed as $T$ is lowered below the Curie temperature. If the conductance is a function of magnetization, this function should be the same underlying the behavior of $\delta G(H,T)$ for $T>T_C$, because the temperature dependence of $G(0,T)$ originates exclusively from the temperature dependence of the  spontaneous magnetization $M(H=0, T)$, which in the ferromagnetic state increases from zero at $T=T_C$, to its saturation value for $T\ll T_C$. Consistently with this idea, the data in Fig.~\ref{fig:tun}e shows an increase in conductance upon lowering $T$. However, to establish whether the functional dependence of the magnetoconductance on magnetization below $T_C$ is the same as the one found for $T>T_C$ a more quantitative analysis is needed.

For such an analysis we cannot rely on the magnetization measured on bulk crystals, because in the ferromagnetic state the magnetization of bulk samples at $H=0$ is entirely determined by the formation of magnetic domains\cite{blundell2003magnetism,kuhlow1975}, whereas exfoliated atomically thin crystals of the size used in our devices have been found to behave as single domains\cite{Kim2019NE,Jin2020NM,Sun2021}. Indeed,  magnetization measurements on CrBr\3 bulk crystals for $T<T_C$ exhibit virtually no remnant magnetization nor any hysteresis upon cycling the applied magnetic field (see Supplementary Fig. 1), whereas Hall magnetometry of atomically thin, exfoliated  CrBr\3 crystals exhibit finite remnant magnetization and a clear hysteresis\cite{Kim2019NE}. That is why in what follows we use the temperature-dependent, zero-field magnetization obtained in Hall magnetometry experiments that --as discussed in Ref.~\onlinecite{Kim2019NE}-- is very well reproduced by the temperature dependence calculated using the XXZ model with anisotropic exchange interaction, shown in Fig.~\ref{fig:scaling}e.

We use the spontaneous magnetization curve shown in Fig.~\ref{fig:scaling}e to re-plot the quantity $\Delta G(H,T)$ shown in Fig.~\ref{fig:scaling}d, as a function of $M$. The result is represented by the open circles in Fig.~\ref{fig:scaling}c. The data fall on top of the $\delta G(M)$ curve found  in our analysis of transport in the paramagnetic state of CrBr\3, for $T>T_C$. The excellent agreement demonstrates that the conductance of CrBr\3  tunnel barriers depends on temperature and magnetic field only through its  magnetization over the full experimental range investigated, and that the dependence is described by the same function from well above $T_C$ to the lowest temperature reached in our measurements (2 K). This conclusion is extremely robust, because it is drawn directly from the analysis of the experimental data, without any theoretical assumption (the 8-layer device exhibits an identical behavior, as discussed in Supplementary Note 3 and shown in Supplementary Fig.~5).

\begin{figure*}
\centering
\includegraphics[width =0.9\linewidth]{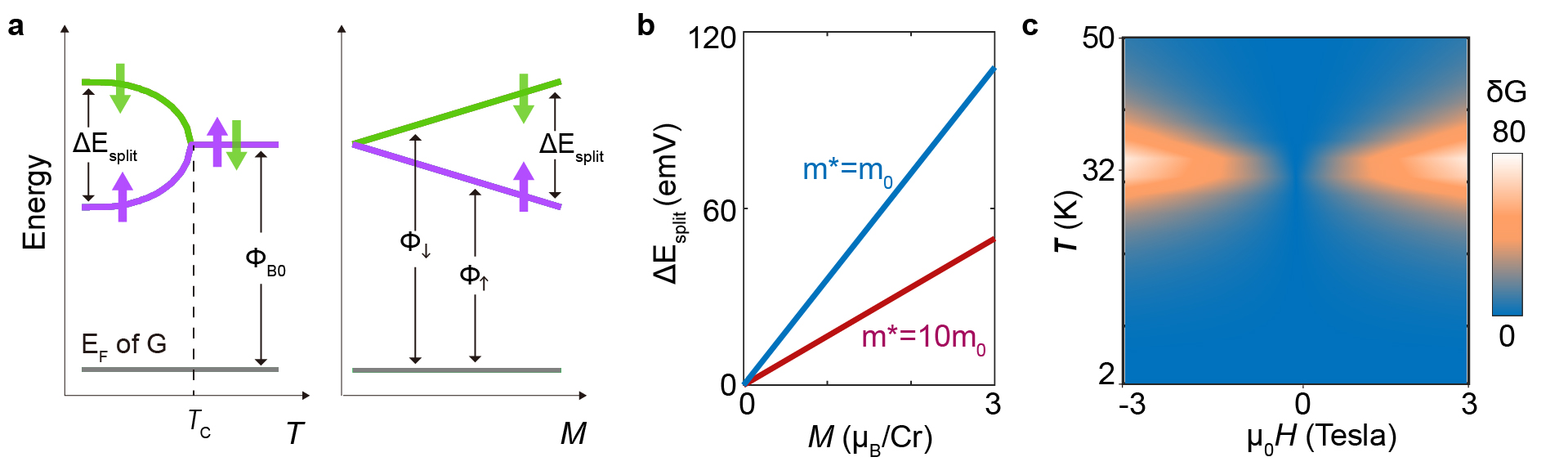}
\caption{\textbf{Relation between spin splitting energy and magnetization.} {\bf a}: Schematic diagram of the relevant energies involved in tunneling process. The spin-up and spin-down CrBr$_3$  conduction bands become non-degenerate when the magnetization $M$ is finite, causing a different height of the tunneling barrier for electrons whose  spin points in opposite directions (the barrier height is given by the distance between the corresponding band edge and the Fermi level $E_F$ in the graphene electrodes, indicated by the bottom horizontal line). A finite $M$ can be induced by the appearance of the spontaneous magnetization as $T$ is lowered below $T_C$, or by the application of an external magnetic field for $T>T_C$. {\bf b}: Magnetization dependence of the splitting energy, as determined from fitting of the data in Fig.~\ref{fig:scaling}{\bf c} using the spin-dependent Fowler-Nordheim tunneling model to model  the conductance. The blue and purple lines correspond to the results obtained assuming the effective mass in the direction perpendicular to the layers of  CrBr$_3$  to be $m_0$ or $10 m_0$, respectively. {\bf c}: Color map of magnetoconductance calculated using Eq. (2) in the main text, using the Weiss model to determine the dependence $M(H,T)$ of the magnetization on magnetic field and temperature.} %
\label{fig:band}
\end{figure*}

These experimental results have a straightforward interpretation within the context of Fowler-Nordheim (FN) tunneling transport commonly used to interpret the conductance of van der Waals magnetic barriers. In FN tunnelling, the applied bias tilts the conduction band across the CrBr\3 layer, effectively reducing the thickness of the tunnel barrier, so that eventually the tunneling  probability for electrons becomes sizable and a finite current is observed\cite{FN1928,FN1969}. The $I-V$ characteristics in the FN tunneling regime satisfy the relation: 
\begin{equation}
   I\propto \frac{V^2}{\phi_B}e^{\frac{-8\pi d \sqrt{2m^*}\phi_B^{3/2} }{3heV}},
\end{equation}
where $m^*$ is the effective mass describing the motion of electrons in CrBr\3 in the direction perpendicular to the planes, $\phi_B$ is the barrier height determined by the distance between the Fermi level in the contacts and the conduction band edge in CrBr\3, $h$ is Planck's constant and $e$ the (modulus of the) electron charge. For a ferromagnet, an analogous relation is expected to hold separately  for electrons with spin up and spin down, which experience different barrier heights $\phi_{\uparrow}$ and $\phi_{\downarrow}$, due to the spin-splitting of the conduction band present for $T<T_C$ (see Fig.~\ref{fig:band}a). The total conductance is then given by the sum  of the contributions given by electrons with spin up and spin down :
\begin{equation}
    G=G^{\uparrow}+G^{\downarrow} = A \frac{V}{\phi_{\uparrow}}e^{\frac{-8\pi d \sqrt{2m^*}\phi_{\uparrow}^{3/2} }{3heV}} + A \frac{V}{\phi_{\downarrow}}e^{\frac{-8\pi d \sqrt{2m^*}\phi_{\downarrow}^{3/2} }{3heV}},
\end{equation}
where $A$ is a constant determined by the barrier dimensions.

We use this expression to analyze the experimental data by assuming  that the spin splitting of the conduction band in the ferromagnetic state is linearly proportional to the magnetization, resulting in barrier heights for spin up and down given by $\phi_{\uparrow,\downarrow}=\phi_{B0} \pm \gamma M$. We calculate the magnetoconductance $[G(M)-G(M=0)]/G(M=0)$ using the value of $\sqrt{2m^*}\phi_{B0}$  extracted from the measured $I-V$ curves, and treating $\gamma$ as the sole fitting parameter. The result of this procedure --represented by the black curve in Fig.~\ref{fig:scaling}c-- reproduces the experimental data perfectly. Interestingly,  a conceptually similar approach has been followed in earlier beautiful work on the tunneling conductance of EuO barriers in the ferromagnetic state\cite{Moodera2008PRL}. That work, however, focused exclusively on the case of  $T<T_C$ and zero applied magnetic field,  by analyzing the temperature dependence of the conductance  in terms of the measured temperature dependence of the magnetization. Our results  show that the approach has a  much broader validity: it can be applied both below and above $T_C$, it remains valid in the presence of a magnetic field, and --as CrBr\3 and EuO are very different materials-- it describes  very different classes of ferromagnetic insulators. If used in conjunction with a model predicting the magnetic field and temperature dependence of the magnetization, this approach  allows the full magnetoconductance to be calculated. This is illustrated by  the color plot in Fig.~\ref{fig:band}d that --despite having being obtained with  the simplest possible Weiss model of Ising ferromagnetism-- reproduces all the qualitative features observed in the experiments (compare Fig.~\ref{fig:band}d with Fig.~\ref{fig:MG}a), and even exhibits a nearly quantitative agreement. Alternatively, it is also possible to extract the temperature and magnetic field dependence of the magnetization from the measured magnetoconductance, as discussed in Supplementary Note 2 and shown in Supplementary Fig.~4.

The excellent agreement between the calculated and the measured magnetoconductance (see Fig.~\ref{fig:scaling}c) suggests the possibility to extract the spin splitting energy quantitatively, from  the value of the fitting parameter $\gamma$. This is however not straightforward, because Eq. (2) depends on the product $\sqrt{m^*}(\phi_0)^{3/2}$, and the effective mass $m^*$ is not known. Fig.~\ref{fig:band}b shows  the spin splitting energy as a function of magnetization obtained by  taking the value of $\gamma$ used to fit the $\delta G = \delta G(M)$ curves in Fig.~\ref{fig:scaling}c, and assuming the effective mass $m^*$ to be either the free electron mass $m_0$ or 10 $m_0$, a very large value chosen to mimic the flatness of the  CrBr\3 bands in the direction perpendicular to the layers\cite{wang2011Calculation,soriano2020review}. We find that at saturation the energy splitting separating the spin-up and the spin-down bands is approximately 110 meV if $m^* = m_0$ and 50 meV if $m^*=10 m_0$, indicating that for realistic values of the effective mass the spin-splitting energy is between 50 and 110 meV. We emphasize, however, that care is certainly needed in interpreting the meaning of this quantity microscopically, because --as applied to CrBr\3 barriers-- the Fowler-Nordheim model is a phenomenological approach that does not  take into account the complexity of the microscopic electronic structure of the material. In particular, it does not take into account that the conduction band consists of two distinct, nearly degenerate electronic bands originating from the $e_{1g}$ and $t_{2g}$ orbitals of the Cr atoms.

Irrespective of these details, the key result presented here is that  the measured tunneling magnetoconductance of CrBr\3 is entirely determined by its magnetization, which is why  magnetoconductance measurements can be used to investigate the magnetic properties of the material. We envision, for instance,  that magnetoconductance measurements will allow detailed investigations of  the critical behavior of the magnetic susceptibility in the paramagnetic state for $T$ very close to $T_C$   and provide a new, experimentally simple way to determine critical exponents. This is possible because  the required data analysis  only relies on the fact that at small $M$ the magnetoconductance is a quadratic function of the magnetization (see Supplementary Note 2 and Supplementary Fig.~3). Another interesting possibility is to analyze magnetoconductance  measurements over a broad range of temperatures and magnetic fields, to discriminate between  microscopic theoretical models that predict  a different functional dependence for $M(H,T)$ (see Supplementary Note 2 and Supplementary Fig.~4). These are just two examples that illustrate the most important aspect of our results, namely that measurements of the tunneling conductance are not limited to the investigation of antiferromagnetic barriers, but can also provide detailed information about the magnetic properties of ferromagnetic insulators.

\bibliographystyle{mynaturemag}
\bibliography{biblio}

\section*{Methods}

\section*{Bulk crystal growth and characterizations}
Crystals of CrBr\3 were grown by the Chemical Vapour Transport method as reported earlier. Pure Chromium (99.95$\%$ CERAC) and TeBr$_4$ (99.9$\%$ Alfa Aesar) were mixed with a molar ratio 1 : 0.75 to a total mass of 0.5 g, and put in quartz tube with an internal diameter of 10 mm and a length of ~13 cm. The preparation of the quartz reactor was done inside a glove box under pure Ar atmosphere. The tube was evacuated down to $~10^{-4}$ mbar and sealed under vacuum, then put in a horizontal tubular furnace in a temperature gradient of about 10°C/cm, with the hot end at 700°C and the cold end at 580°C. After 7 days at this temperature, the furnace was switched off, and the tube cooled to room temperature. CrBr\3 was found to crystallise at the cold end of the tube. Shiny, thin platelet-like, dark green-blackish single crystals of typical lateral size of 2-5 mm were extracted. Bulk crystals of ~2.1 mg were used for the magnetic characterization in a MPMS3 SQUID magnetometer (Quantum Design). The magnetic moment of the crystals was measured with magnetic field parallel to the crystallographic c-axis. 

\subsection*{Tunneling junction fabrication and transport measurements}
CrBr\3 multilayers were mechanically exfoliated from the crystals discussed in the section of crystal growth. Tunnel junctions of multilayer graphene/CrBr\3/multilayer graphene were assembled using a pick-and-lift technique with stamps of PDMS/PC. To avoid degradation of thin CrBr\3 multilayers, the exfoliation of CrBr\3 and the heterostructure stacking process were done in a glove box filled with Nitrogen gas, and the whole tunneling junction was encapsulated with hBN before being taken out. Conventional electron beam lithography, reactive-ion etching, electron-beam evaporation (10 nm/50 nm Cr/Ar) and lift-off process were used to make edge contacts to the multilayer graphene. The thickness of the layers was determined by atomic force microscope measurements performed outside the glove box, on the encapsulated devices. Transport measurements were performed in a cryostat from Oxford Instruments, using home-made low-noise electronics. 

\section*{Data availability}
All relevant data are available from the corresponding authors upon reasonable and well-motivated request.

\section*{Acknowledgements}
We sincerely acknowledge Alexandre Ferreira for technical support. Z.W. acknowledges the National Natural Science Foundation of China (Grants no. 11904276). A.F.M. gratefully acknowledges financial support from the Swiss National Science Foundation (Division II) and from the EU Graphene Flagship project. M.G.\ acknowledges support  from the Italian Ministry for University and Research through the Levi-Montalcini program. K.W. and T.T. acknowledge support from the Elemental Strategy Initiative conducted by the MEXT, Japan ,Grant Number JPMXP0112101001 and JSPS KAKENHI Grant Number JP20H00354.

\section*{Author contributions}
Z.W.\ and A.F.M.\ conceived the work. D.D.\ and E.G.\ grew CrBr\3 crystals and performed bulk characterization. T.T.\ and K.W.\ provided high-quality boron nitride crystals. Z.W.\ fabricated samples and performed transport measurements with help of I.G.\ and N.U.. Z.W.,  I.G., N.U., M.G.\ and A.F.M.\ analyzed and interpreted the magnetoconductance data. All authors contributed to writing the manuscript.

\section*{Competing interests}
The authors declare no competing interests.

\end{document}